\documentclass[12pt]{iopart}
\usepackage{amssymb}
\usepackage{graphicx,bm, cite,color}
\usepackage{setstack}
\usepackage{lscape}
\begin{document}
	
	\title[]{Quantum solvability of quadratic Li\'{e}nard type nonlinear oscillators possessing maximal Lie point symmetries: An implication of arbitrariness of ordering parameters}
	\author{V. Chithiika Ruby and M. Lakshmanan}
\address{Center for Nonlinear Dynamics, School of Physics,
		Bharathidasan University, Tiruchirapalli - 620 024, India.}
	
\begin{abstract}
In this paper, we investigate the quantum dynamics of underlying two one-dimensional quadratic Li\'{e}nard type nonlinear oscillators which are classified under the category of maximal (eight parameter) Lie point symmetry group (J. Math. Phys. 54 , 053506 (2013)). Classically, both the systems were also shown to be linearizable as well as isochronic. In this work, we study the quantum dynamics of the nonlinear oscillators by considering a general ordered position dependent mass Hamiltonian. The ordering parameters of the mass term are treated to be arbitrary to start with. We observe that the quantum version of these nonlinear oscillators are exactly solvable provided that the ordering parameters of the mass term are subjected to certain constraints imposed on the arbitrariness of the ordering parameters. We obtain the eigenvalues and eigenfunctions associated with both the systems.  We also consider briefly the quantum versions of other examples of quadratic Li\'{e}nard oscillators which are classically linearizable.
\end{abstract}

\maketitle

\section{Introduction}
Classically certain Li\'{e}nard type-I and type-II nonlinear oscillators were shown to possess nonstandard Hamiltonians characterized by the nonlinear parameter \cite{mathews-lakshmanan,higgs,carinena2004non,ball,chandru,tiwari,omustafa}. Under appropriate limit on the system parameters or through  suitable transformations (local/ nonlocal), the Hamiltonians can be related to that of the linear harmonic oscillator. The Hamiltonians are observed to be position dependent mass Hamiltonians. In the literature many authors studied the isochronous/ nonisochronous, linearizable and integrable nature of these nonlinear oscillators. For example, the nonlinear oscillators studied by Mathews and Lakshmanan and Higgs were shown to be exactly solvable both at classical and quantum regimes \cite{mathews-lakshmanan,higgs,carinena2004non}. They possess non-isochronous solutions at the classical level and admit nonlinear energy spectrum in the  quantum regime in which the nonlinear parameter appears in the energy spectrum. On the other hand, the nonlinear oscillator characterized by the modified Emden equation \cite{chandru}  and a class of nonlinear systems studied by  Tiwari etal. admit isochronous solutions  \cite{tiwari}. The quantum counterpart of the  associated time independent classical Hamiltonian of the former system has been shown to be {\it PT}-invariant (parity-time invariant) and  the underlying time independent Schr\"{o}dinger equation has been solved in momentum space which yields a linear energy spectrum \cite{pt-chithiika}. We infer from the various studies on these types of  Li\'{e}nard type nonlinear oscillators possessing position dependent mass Hamiltonians that the nonlinear oscillators can be grouped into (i) Non-isochronous position dependent mass oscillators and (ii) isochronous position dependent mass oscillators. 

In this paper, we consider the quantum dynamics of quadratic Li\'enard type nonlinear oscillators possessing isochronous solutions. Interestingly, while studying the quantum counterpart of the quadratic Li\'enard type nonlinear oscillators admitting non-isochronous periodic oscillations, it has been realized to possess nonlinear energy spectrum characterized by the nonlinear parameter which also determines the range of coordinate space about which the periodic solutions are admitted. In the absence of the nonlinear parameter, the systems match with that of the quantum harmonic oscillator \cite{mathews-lakshmanan,higgs,carinena2004non,karthiga}. In contrast,  the quadratic Li\'enard type nonlinear oscillators which admit isochronous periodic oscillations within the range of coordinate space defined by the nonlinear parameter generally seem to possess a linear energy spectrum in the quantum regime, which  is free from the nonlinear parameter. Hence, they may be called  ``isotonic oscillators" as they match with that of linear harmonic oscillator. In this paper, we study the  isochronicity nature of the quadratic Li\'{e}nard type nonlinear oscillators and investigate how it is preserved in the quantum regime with specific examples in this work.  

Recently, Tiwari et al. proposed a systematic technique to identify the quadratic Li\'{e}nard type second order nonlinear differential equations which admit one, two, three and eight parameter Lie point symmetry groups and classified the equations as well \cite{tiwari}. They have shown that the quadratic Li\'enard oscillators possessing eight parameter (maximal) Lie point symmetries are linearizable and isochronous and obtained general conditions for isochronicity (see Appendix A for brief details). They also discussed two specific examples of nonlinear oscillators which are linearizable by admitting the maximal (eight parameter) symmetry group and the corresponding quadratic Li\'{e}nard type equations are written as, 
\begin{eqnarray}
\ddot{x} + \lambda \dot{x}^2 + \frac{\omega^2_0}{\lambda} \left(1 - e^{-\lambda\;x}\right) &=& 0, 
\label{lie1}
\end{eqnarray}
and
\begin{eqnarray}
\ddot{x} - \frac{2\;\lambda}{1 + \lambda x} \dot{x}^2 +\omega^2_0 \left(x + \lambda x^2\right) &=& 0,  
\label{lie2}
\end{eqnarray}
whose Lagrangians can be written as 
\begin{eqnarray}
L_1 &=& \frac{\lambda^2\;e^{2\;\lambda\;x}}{2\; }\;\dot{x}^2 - \frac{\omega^2_0}{2}\left(1 - e^{\lambda\;x}\right)^2,\label{Lag1}\\
L_2 &=& \frac{\dot{x}^2}{2\;(1 + \lambda x)^4 } - \frac{\omega^2_0\; x^2}{2\;(1 + \lambda x)^2} . \label{Lag2}
\end{eqnarray}
The corresponding Hamiltonians are obtained, respectively, as 
\begin{eqnarray}
H_1 &=& \frac{1}{2\; \lambda^2}\;e^{-2\;\lambda\;x} p^2 + \frac{\omega^2_0}{2}\left(1 - e^{\lambda\;x}\right)^2, \qquad p = {\lambda^2}\; e^{2\;\lambda\;x}\;\dot{x}, \label{ham1}\\
H_2 &=& \frac{1}{2}(1 + \lambda x)^4 p^2 + \frac{\omega^2_0\; x^2}{2\;(1 + \lambda x)^2},\qquad p = \frac{\dot{x}}{(1+\lambda x)^4}. \label{ham2}
\end{eqnarray}
We will consider the quantum treatment for the above two equations here. Other examples (see for example, \cite{omustafa}) can be treated in a similar fashion as these examples as indicated in Appendix B. 
In Eqs. (\ref{ham1}) and (\ref{ham2}), $p$ is the canonically conjugate momentum associated with the position variable $x(t)$.  
In this work, we wish to consider the corresponding quantum versions of the systems which naturally become quantum systems with position-dependent effective mass forms. Position-dependent mass (PDM) quantum systems find  applications in condensed matter physics \cite{apps}, quantum dots and quantum wells  \cite{qdots, qwell} and so on. Classically the systems (\ref{ham1}) and (\ref{ham2}) can be solved exactly and they have been shown to possess isochronous solutions \cite{tiwari}, see Appendix for some details.  While quantizing, the position dependent mass systems require appropriate ordering between momentum and mass operators in the kinetic energy term and also require appropriate modifications in the boundary conditions since some mass functions may not be continuous \cite{bou}. In the literature, many different orderings are available such as Weyl ordering \cite{weyl_pdms, weyl_ord}, von Roos ordering \cite{von}, Li and Kuhn ordering \cite{lik} and Zhu and Kroemer  ordering \cite{zhu}, etc.  which correspond to the Hermitian construction of the associated quantum Hamiltonian. Recently, Trabelsi etal. \cite{Trabelsi}  proposed a general formulation of the kinetic energy operator with PDM under which von Roos ordering comes as one of the possibilities. In ref. \cite{chithiika-removal}, we considered the corresponding generalized $2 N$-parameter kinetic energy operator which  unifies all types of Hermitian and non-Hermitian orderings and investigated the effect of ambiguity in the dynamics of systems endowed with one-dimensional potentials. The associated Hamiltonian of  the generalized  kinetic energy operator is as such non-Hermitian and can become Hermitian on either applying a specific condition on the ordering parameters or through a similarity transformation. We also observed that a class of non-hermitian ordered Hamiltonian systems can be shown to be exactly solvable because of their quasi-Hermitian property. However, we cannot conclude that the quantum systems are exactly solvable  for arbitrary choices of ordering parameters. 

The solvability of position dependent mass systems has also been studied using traditional Hartree second order shooting method \cite{kill}, point canonical transformation method \cite{pct}, Lie algebraic approach \cite{Roy} and  from the point of view of PT-symmetry \cite{apzhang}. 
Besides these studies on the quantum solvability of position dependent mass systems in connection with ordering ambiguity, in the literature different algebraic techniques using  the superintegrability of nonlinear systems are also available for higher dimensional position dependent mass systems. Examples for such studies are the two dimensional Higgs oscillator and superintegrable generalizations of Higgs oscillator \cite{Bonatas, ball1, ball2, ball3}.

In this paper, we have considered the $2N$-parameters generalized Schr\"{o}dinger equation in order to study the quantum dynamics of the two systems (\ref{ham1}) and (\ref{ham2}). We observe that the two systems are exactly solvable for certain choices of ordering parameters and their energy eigenvalues are isotonic as that of the linear harmonic oscillator. Hence, that isochronous property is also preserved in the quantum regime as well for appropriate ordering.  The plan of the paper is as follows. In section \ref{sec1}, we discuss the quantum solvability of nonlinear oscillators by considering the general form of PDM kinetic energy operator. We consider the two different forms of the nonlinear oscillators (\ref{ham1}) and (\ref{ham2}) and discuss their quantum solvability in sections \ref{hamiltonian1} and \ref{hamiltonian2}, respectively. In Sec. 5, we consider the non-Hermitian ordering of the systems and discuss their quantum solvability. Finally, in Sec. \ref{conc}, we summarize our results. In the Appendix A, we summarize the results on the classical dynamics of the nonlinear oscillators (\ref{ham1}) and (\ref{ham2}) and discuss about the quantum solvability of other examples of quadratic Li\'{e}nard type nonlinear oscillators possessing isochronous solutions in Appendix B.

\section{\label{sec1} Quantum solvability of nonlinear oscillators}
In general the Hamiltonians (\ref{ham1}) and (\ref{ham2}) are of the form 
\begin{equation}
H = \frac{p^2}{2\;m(x)} + V(x), \label{pdm}
\end{equation}
where $m(x)$ is the position dependent mass term. 

To study the quantum solvability of the system (\ref{pdm}), we consider the most general form of associated Hamiltonian operator that provides a complete classification of Hermitian and non-Hermitian orderings \cite{Trabelsi}, 
\begin{eqnarray}
\hat{H} = \frac{1}{2}\sum^N_{i = 1} w_i m^{\alpha_i} \hat{p} m^{\beta_i} \hat{p} m^{\gamma_i} + V(x), 
\label{geo}
\end{eqnarray}
where  $N$ is an arbitrary positive integer and $\hat{p}$ is the one dimensional momentum operator. The ordering parameters should satisfy 
the constraints, $\alpha_i +\beta_i +\gamma_i = -1,\;i =1, 2, 3, ... N,$ and $w_i$'s  are real weights which are summed to be $1$. The above form globally connects all the Hermitian orderings and also provides a complete classification of Hermitian and non-Hermitian orderings \cite{Trabelsi}.  The operator $\hat{H}$ in (\ref{geo}) possesses $2 N$ free ordering parameters, after taking into account the above constraints.

The corresponding Hamiltonian  for the potential $V$ can be written as  
\begin{equation}
\hspace{-0.5cm} \hat{H}_{non} = \frac{1}{2}\hat{p}\frac{1}{m}\hat{p}+(\bar{\gamma} - \bar{\alpha}) \frac{i \hbar}{2} {\frac{d}{dx}} \left( \frac{1}{m}\right)\hat{p} + \frac{\hbar^2}{2}\left[\bar{\gamma} {\frac{d^2}{dx^2}}\left(\frac{1}{m}\right) + \overline{\alpha\gamma} \left(\frac{m'^2}{m^3}\right) \right] + V, \label{nhe}
\end{equation}
where ${\displaystyle \hat{p} = -i\hbar \frac{d}{dx}}$ and the subscript $non$ in $\hat{H}_{non}$ implies that the Hamiltonian is non-Hermitian. In (\ref{nhe}), the over bar over the parameters represent their total value, $\bar{X} = \sum^{N}_{i} w_i X_i$.   

The non-Hermitian Hamiltonian ${ \hat{H}_{non}}$ given by (\ref{nhe}) can be related to the Hermitian Hamiltonian ${\hat{H}_{her}}$  by performing the similarity  transformation 
\begin{equation}
\hat{H}_{her}  = m^{\eta} {\hat{H}}_{non} m^{-\eta}, \qquad 2 \eta = \bar{\gamma} - \bar{\alpha},  \label{heta} 
\end{equation}
which yields
\begin{eqnarray}
\hat{H}_{her} &=& \frac{1}{2}\hat{p}\frac{1}{m}\hat{p}+ \frac{\hbar^2}{2}\left[\bar{\gamma} {\frac{d^2}{dx^2}}\left(\frac{1}{m}\right) + \overline{\alpha\gamma} \left(\frac{m'^2}{m^3}\right) \right] + V.  
\label{geham_he1}
\end{eqnarray}

The time-independent Schr\"{o}dinger equation for the Hamiltonian (\ref{geham_he1}), $\hat{H}_{her}\psi = E \psi,$ can be written as   
\begin{eqnarray}
\hspace{-2.5cm} \psi{''} - \frac{m{'}}{m} \psi{'} + 
\left(\left(\frac{\bar{\alpha}+\bar{\gamma}}{2}\right)  \frac{m{''}}{m}-\left(\overline{\alpha\gamma}+\bar{\gamma}+\bar{\alpha}+\frac{1}{4}(\bar{\gamma} - \bar{\alpha})^2\right)  \frac{m{'^2}}{m^2}\right) \psi + \frac{2 m}{\hbar^2}\left(E -V(x)\right)\psi  = 0, \nonumber \\
\label{seg_he}
\end{eqnarray}
where ${\displaystyle ' = \frac{d}{dx}}$. Since the above Hamiltonian depicts the dynamics of one dimensional potentials (\ref{ham1}) and (\ref{ham2}) also, we use the generalized position dependent mass Schr\"{o}dinger equation (\ref{seg_he}) concerning the Hermitian ordering to study the solvability of the systems in the following sections.
 
\section{\label{hamiltonian1} Nonlinear oscillator- Exponential form}
To start with let us consider the nonlinear oscillator (\ref{ham1}) with the exponential form allowing eight parameter Lie point symmetries, 
\begin{equation}
H_1 = \frac{1}{2\; \lambda^2}\;e^{-2\;\lambda\;x} p^2 + \frac{\omega^2_0}{2}\left(1 - e^{\lambda\;x}\right)^2,\nonumber 
\end{equation}
where the mass term and the potential are expressed as
\begin{equation}
m(x) = \lambda^2 e^{2\;\lambda\;x}, \qquad V_1(x) = \frac{\omega^2_0}{2}\left(1 - e^{\lambda\;x}\right)^2. \label{mass1}
\end{equation}
The generalized Schr\"{o}dinger equation (\ref{seg_he}) now becomes 
\begin{equation}
\psi'' - 2 \lambda \psi' +  \lambda^2 \left[A + \xi e^{2\;\lambda\; x} - \mu^2 (e^{2\;\lambda\;x} - 2 e^{3\lambda\;x} + e^{4\;\lambda\;x})\right] \psi = 0, \label{geq1}
\end{equation}
where the terms $A,\; \xi, \mu$ are defined as, 
\begin{eqnarray}
  A & = & - 4 {\overline{\alpha \gamma}} - (\bar{\gamma} - \bar{\alpha})^2 - 2 (\bar{\gamma} + \bar{\alpha}), \label{termA}\\
\xi & = & \frac{2\;E}{\hbar^2}, \\
\mu & = & \frac{\omega_0}{\hbar}. 
\end{eqnarray}
On using the following transformations, 
\begin{eqnarray}
z = e^{\lambda\; x}, \qquad \psi(z) = \exp{\left(-\frac{\mu}{2}z^2 +\mu z\right)}\; z^{d} \phi(z), 
\label{trans1}
\end{eqnarray}
we can reduce the equation (\ref{geq1}) to the form
\begin{eqnarray}
\hspace{-3cm} z^2\;\phi''(z) + (2d-1 + 2 \mu z - 2 \mu z^2)\;z\;\phi'(z) +\left[A + d (d-2) + \mu (2d-1) z + (\xi - 2 \mu d) z^2 \right] \phi(z) = 0, \nonumber \\ \label{geqf}
\end{eqnarray}
where ${\displaystyle ' = \frac{d}{dz}}$. 
Here, in Eq. (\ref{geq1}), the parameter $\lambda$ can be of either sign, that is positive or negative. If $\lambda > 0$, the variable $z \in (0, \infty)$ for $-\infty < x < \infty$. And for $\lambda < 0$, the variable $z \in (\infty, 0)$ for $-\infty < x < \infty$. Hence we solve the equation (\ref{geq1}) for $\lambda > 0$ and then express the solution for both the cases. 

\subsubsection{Case $\lambda >0$}
By fixing the values of $d$ and ordering term $A$ as follows, 
\begin{eqnarray}
d &=& \frac{1}{2},\label{dvalue}\\ 
A &=& \frac{3}{4} \rightarrow 4 {\overline{\alpha\gamma}} + (\bar{\gamma} - \bar{\alpha})^2 + 2 (\bar{\gamma} + \bar{\alpha}) = -\frac{3}{4}, \label{avalue}
\end{eqnarray}
and using the transformation $\tau = \sqrt{\mu}\;(z-1)$, we can reduce the equation (\ref{geqf}) to the form 
\begin{equation}
\phi''(\tau) - 2 \tau\;\phi'(\tau) + \left(\frac{\xi}{\mu} - 1\right) \phi(\tau) = 0. \label{geq2se}
\end{equation}
Eq. (\ref{geq2se}) is of the form of the Hermite differential equation, $H^{''}_n(x) - 2 x H^{'}_n(x) + 2 n = 0$. Using the equation (\ref{trans1}), we can write down the eigenfunctions of (\ref{geq1}) as, 
\begin{equation}
\hspace{-2.5cm} \psi_n(x) = N_n\;\exp{\left(-\frac{\omega_0}{2\;\hbar}e^{2\;\lambda x} + \frac{\omega_0}{\hbar} e^{\lambda x}\right)}\; e^{\lambda x/2} \; 
H_n\left[\sqrt{\frac{\omega_0}{\hbar}}\;\left(e^{\lambda\;x}-1\right)\right], \quad -\infty < x < \infty, \label{eigen1}
\end{equation}
with energy eigenvalues $E_n$ as
\begin{equation}
 E_n = \left(n + \frac{1}{2}\right)\;\hbar\;\omega_0,\qquad n = 0, 1, 2, 3, ...,  \label{en1}
\end{equation}
where $N_n, \; n = 0, 1, 2, 3, ...,$ is the normalization constant. The value of $N_n$ can be obtained from the relation
\begin{eqnarray}
1 & = & \langle \psi_n | \psi_n\rangle \nonumber \\
  & = & 2 N^2_n \int^{\infty}_0 \exp{\left(-\frac{\omega_0}{\;\hbar}e^{2\;\lambda x} + 2\frac{\omega_0}{\hbar} e^{\lambda x}\right)}\; e^{\lambda x}\nonumber \\ & & \times H_n\left[\sqrt{\frac{\omega_0}{\hbar}}\;\left(e^{\lambda\;x}-1\right)\right]H_n\left[\sqrt{\frac{\omega_0}{\hbar}}\;\left(e^{\lambda\;x}-1\right)\right] dx, \nonumber
\end{eqnarray}
which can be evaluated to obtain the normalization constant as 
\begin{eqnarray}
N_n = \left(e^{-\frac{\omega_0}{\hbar}}\;\frac{\sqrt{\frac{\omega_0}{\hbar}}\;\lambda}{\frac{\sqrt{ \pi}}{2}\;2^{n}\;n!\left(1+erf(a)\right)+a e^{-a^2}O(a^2)}\right)^{1/2}, \label{nn1} 
\end{eqnarray}
where $erf(a)$ is the error function and $a = \sqrt{\frac{\omega_0}{\hbar}}.$

\subsubsection{Case $\lambda < 0$}
As we pointed out earlier, we can express the solution in terms of negative values of $\lambda$ by simply replacing $\lambda = - |\lambda|$ in (\ref{eigen1})
\begin{eqnarray}
\hspace{-2.5cm} \psi_n(x) = N_n\;\exp{\left(-\frac{\omega_0}{2\;\hbar}e^{-2\;|\lambda| x} + \frac{\omega_0}{\hbar} e^{-|\lambda| x}\right)}\; e^{-|\lambda| x/2} \; 
H_n\left[\sqrt{\frac{\omega_0}{\hbar}}\;\left(e^{-|\lambda|\;x}-1\right)\right],  -\infty < x < \infty,\nonumber \\
\end{eqnarray}
with energy eigenvalues, $E_n$ as
\begin{equation}
 E_n = \left(n + \frac{1}{2}\right)\;\hbar\;\omega_0, \qquad n = 0, 1, 2, 3, ...,
 \label{en2}
\end{equation}
where, $N_n, \; n = 0, 1, 2, 3, ...,$ is normalization constant,which can be evaluated as   
\begin{eqnarray}
N_n = \left(e^{-\frac{\omega_0}{\hbar}}\;\frac{\sqrt{\frac{\omega_0}{\hbar}}\;|\lambda|}{\frac{\sqrt{ \pi}}{2}\;2^{n}\;n!\left(1+erf(a)\right)+a\; e^{-a^2}O(a^2)}\right)^{1/2}, \label{nn2} 
\end{eqnarray}
where $erf(a)$ is the error function and $a = \sqrt{\frac{\omega_0}{\hbar}}.$

We observe that the energy eigenvalues (vide (\ref{en1}) and (\ref{en2})) are linear in quantum number $n$ and independent of both the nonlinear parameter $\lambda$ and the ordering parameters. Hence we conclude that the isochronous property of the classical system (\ref{ham1}) is preserved in its corresponding  quantum counterparts also.

\begin{figure}[!ht]
\vspace{0.5cm}
\begin{center}
\includegraphics[width=0.5\linewidth]{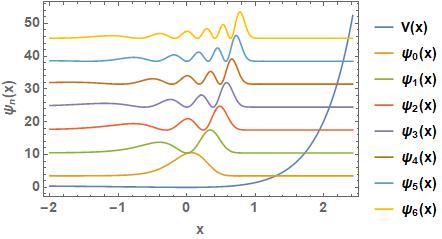}
\end{center}
\vspace{-0.2cm}
\caption{The plot of $V(x)$ along with the eigenfunctions $\psi_n(x)$ for $n=0,1,2,3,4,5$ and $6$ for $\lambda =1$ and $\omega_0 = 7$.} \label{pot1}
\vspace{-0.3cm}
\end{figure}

\begin{figure}[!tbp]
  \centering
  \begin{minipage}[b]{0.4\textwidth}
    \includegraphics[width=\textwidth]{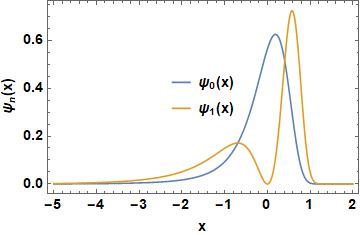}
     \end{minipage}
  \hfill
  \begin{minipage}[b]{0.4\textwidth}
    \includegraphics[width=\textwidth]{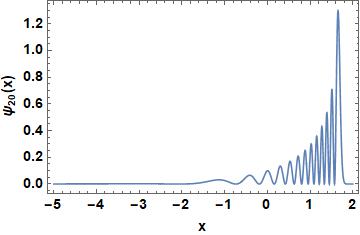}
  \end{minipage}
  \vspace{-0.2cm}
  \caption{The plot of $\psi_0(x)$, $\psi_1(x)$ and $\psi_{20}(x)$ for $\lambda =1$ and $\omega_0 = 2$.} \label{eig1}
  \vspace{0.3cm}
\end{figure}

Both the figures \ref{pot1} and \ref{eig1} are plotted for positive value of $\lambda$. From these plots, we infer that in the limit $x$ tends to $+\infty$, the potential $V_1(x)$ characterized by $e^{2 \lambda x}$ grows to infinite value exponentially and so the probability of finding the particle is maximum near the potential and $|\psi_n(x)|^2$ becomes zero at $x = +\infty$, whereas $V_1(x)$ approximately takes a constant value $\frac{\omega_0^2}{2}$ in the range of $x\in(-\infty, \ln(2)/\lambda)$, where the eigenfunction becomes zero. It mimics the harmonic oscillator spectrum when its quantum oscillations are restricted within the positive half-range of the coordinate space. This is due to the fact that the mass is a varying one with position.  

Hence, with the choice of the values of $d$ and $A$ (vide (\ref{dvalue}) and (\ref{avalue})), the system is found to be exactly solvable. For other choice of $A = -d (d-2)$, equation (\ref{geqf}) is reduced to the form  
\begin{equation}
\hspace{-2cm}\phi''(z) + \left[\frac{2d-1}{z} + 2 \mu  - 2 \mu z\right]\;\phi'(z) +\left[\frac{\mu (2d-1)}{z}  + (\xi - 2 \mu d)  \right] \phi(z) = 0, \label{bich}
\end{equation}
where ${\displaystyle ' = \frac{d}{dz}}$. Eq. (\ref{bich}) is of the form of the bi-confluent Heun equation which might be quasi-exactly solvable \cite{heun}.  

\subsection{Ordering parameters and quantum exactly solvability}
To illustrate the role of ordering parameters in the case of exactly solvable examples of the system (\ref{ham1}), we consider two different ordering forms such as (i) von Roos ordering \cite{von} and (ii) the symmetric ordered form proposed by Gora and Williams \cite{qdots}. The von Roos ordering, 
\begin{eqnarray}
\hat{H}_1 = \frac{1}{4}\left[m^{\alpha}\;\hat{p}\;m^{\beta}\;\hat{p}\;m^{\gamma} + m^{\gamma}\;\hat{p}\;m^{\beta}\;\hat{p}\;m^{\alpha}\right] + V(x), 
\end{eqnarray}      
 is obtained from (\ref{geo}) for the values, $\alpha_1 = \alpha,\;\beta_1=\beta,\;\gamma_1=\gamma$ and $\alpha_2 = \gamma,\;\beta_2=\beta,\;\gamma_2=\alpha,$ with the weights $w_1=w_2=\frac{1}{2}.$ We can evaluate their mean values, 
\begin{equation}
\bar{\alpha} = \frac{\alpha+\gamma}{2}, \quad \bar{\beta} = \beta, \quad \bar{\gamma} = \frac{\gamma+\alpha}{2}\qquad \mbox{and}\qquad\bar{\alpha\gamma}= \alpha\gamma.
\end{equation}
Equation (\ref{avalue}) and the condition $\bar{\alpha}+\bar{\beta}+\bar{\gamma}=-1$  become 
\begin{equation}
2\alpha\gamma + \alpha + \gamma = -\frac{3}{8}, \quad \mbox{and}\quad \alpha + \beta + \gamma = -1.
\end{equation}
Here, the ordering parameters $\alpha, \beta$ and $\gamma$ are arbitrary. It is proved that the von Roos ordered form of the position dependent mass system (\ref{ham1}) is exactly solvable. 

Secondly, we consider the Gora and Williams form, 
\begin{eqnarray}
\hat{H}_1 = \frac{1}{2}\left[\frac{1}{2\;m}\hat{p}^{2}+ \hat{p}^2\;\frac{1}{2m}\right] + V(x). \label{gwform}
\end{eqnarray}
It can be derived from (\ref{geo}) for the choices $\alpha_1=-1 =\gamma_2, \; \alpha_2=\beta_1 = \beta_2 = \gamma_1 = 0$ with weights ${\displaystyle w_1 = w_2 = \frac{1}{2}.}$  One can easily verify that the corresponding mean values, $\bar{\alpha}=\bar{\gamma} =\frac{-1}{2},\; \bar{\beta} = 0$ and $\bar{\alpha\gamma}=0$, do not satisfy the condition (\ref{avalue}). Hence, the position dependent mass system (\ref{ham1}) corresponding to the ordering form (\ref{gwform}) is not exactly solvable. In general, the advantage of using the general ordered form (\ref{geo}) helps one to find out a class of exactly solvable quantum systems for the given mass profile and the potential. Hence, different choices of the ordering parameters subjected to the constraint (\ref{avalue}) gives different types of exactly solvable  position dependent mass systems (\ref{ham1}) which all admit the same set of eigenvalues and eigenfunctions. 

\section{\label{hamiltonian2} Nonpolynomial momentum dependent oscillator}
Let us next consider the nonpolynomial nonlinear oscillator (\ref{ham2}) allowing eight parameter Lie point symmetries, 
\begin{equation}
\hspace{0.7cm}H_2 = \frac{1}{2}\;(1 + \lambda x)^4 p^2 + \frac{\omega^2_0}{2}\;\frac{x^2}{(1 + \lambda x)^2},\nonumber 
\end{equation}
where the mass term and the potential are identified as
\begin{equation}
m(x) = \frac{1}{(1 + \lambda x)^4}, \qquad V_2(x) = \frac{\omega^2_0}{2}\;\frac{x^2}{(1 + \lambda x)^2}. \label{mass2}
\end{equation}
The generalized Schr\"{o}dinger equation (\ref{seg_he}) now becomes 
\begin{equation}
\hspace{-0.5cm} \psi'' + \frac{4\;\lambda}{1 + \lambda x} \psi' +  \left[\frac{B\; \lambda^2}{(1 + \lambda x)^2} + \frac{2\;E}{\hbar^2\;(1 + \lambda x)^4} - \frac{\omega^2_0\;x^2}{\hbar^2\;(1 + \lambda\;x)^6}\right] \psi = 0, \label{geq2}
\end{equation}
where the term $B$ is defined as 
\begin{eqnarray}
  B & = & - 16 {\overline{\alpha \gamma}} - 4\;(\bar{\gamma} - \bar{\alpha})^2 - 6 (\bar{\gamma} + \bar{\alpha}). \label{termB}
\end{eqnarray}
By using the transformations, 
\begin{eqnarray}
z = \frac{1}{1 + \lambda x}, \qquad  \psi(z) = e^{\displaystyle{-\frac{\mu}{2}(1-z)^2}}\;z^d\;\phi(z),
\label{trans2}
\end{eqnarray}
with ${\displaystyle \mu = \frac{\omega_0}{\hbar\;\lambda^2}}$, 
we can reduce equation (\ref{geq2}) to the form 
\begin{eqnarray}
\hspace{-1.5cm} z^2\;\phi''(z) +\left[2(d-1)z + 2\;\mu\;z^2(1 - z)\right]\phi'(z) +\left[d(d-3)+B + 2\;\mu\;(d-1)\;z \right. \nonumber \\
\hspace{-1.5cm} \left.\hspace{3cm}+ \left(\frac{2\;E}{\hbar^2\;\lambda^2} + \mu - 2\mu d \right) z^2 \right] \phi(z) = 0, \label{geq3}
\end{eqnarray}
where ${\displaystyle ' = \frac{d}{dz}}$. 
Here, when $\lambda > 0$, $z \in (\infty, 0)$ for $x \in \left(-\frac{1}{\lambda}, \pm \infty\right)$, whereas for $\lambda < 0$, $z \in (\infty, 0)$ for $x \in \left(\frac{1}{|\lambda|}, \pm \infty\right)$. So we solve equation (\ref{geq3}) for $\lambda > 0$ and then we write the solution for (\ref{geq3}) when $\lambda < 0$ by simply replacing $\lambda = -|\lambda|$. 

\subsubsection{Case $\lambda > 0$}
With the constraints imposed on the ordering parameters, 
\begin{equation}
 d =1, \qquad \mbox{and}\qquad B = 2 \rightarrow 8 {\overline{\alpha \gamma}} + 2\;(\bar{\gamma} - \bar{\alpha})^2 + 3 (\bar{\gamma} + \bar{\alpha}) = -1,   \label{constraint2}
\end{equation} 
and using the transformation, 
\begin{equation}
\tau = (1-z)\;\sqrt{\mu} \qquad \mbox{and} \qquad d =1, \qquad  
\end{equation} 
we can reduce the equation (\ref{geq3}) to  
\begin{equation}
\phi''(\tau) - 2 \tau\;\phi'(\tau) + \left(\frac{2\;E}{\hbar^2\;\lambda^2\;\mu} - 1\right)\;\phi(\tau) = 0. \label{geq4}
\end{equation}
It is again of the form of Hermite differential equation, with the identification $\phi(\tau) = H_n(\tau)$. Then, we can write down the eigenfunctions by resubstituting  the transformations, 
\begin{equation}
\hspace{-2.5cm} \psi_n(x) = \left\{
\begin{array}{ccc}
& & \hspace{-0.6cm} N_n\;\frac{1}{(1 + \lambda x)}\;\exp{\left(-\frac{\omega_0\;x^2}{2\;\hbar\;(1 + \lambda x)^2}\right)}\; 
H_n\left[\sqrt{\frac{\omega_0}{\hbar}}\;\left(\frac{x}{1 + \lambda x}\right)\right], \quad -\frac{1}{\lambda} < x < \infty,\nonumber \\
          &  &\hspace{-0.6cm}0, \hspace{8.5cm}\quad x < -\frac{1}{\lambda}, \label{eigen3} 
\end{array}\right. 
\end{equation}
with energy eigenvalues, $E_n$ as
\begin{equation}
E_n = \left(n + \frac{1}{2}\right)\;\hbar\;\omega_0,\qquad n = 0, 1, 2, 3, ..., \label{en3}
\end{equation}
where $N_n, \; n = 0, 1, 2, 3, ...,$ is the normalization constant, which can be obtained from 
\begin{eqnarray}
 1 & = & \langle \psi_n | \psi_n\rangle \nonumber \\
  & = &  N^2_n \int^{\infty}_{-1/\lambda}\frac{1}{(1 + \lambda x)^2}\;\exp{\left(-\frac{\omega_0\;x^2}{\hbar\;(1 + \lambda x)^2}\right)}\; 
 H_n\left[\sqrt{\frac{\omega_0}{\hbar}}\;\left(\frac{x}{1 + \lambda x}\right)\right]\nonumber\\
 & & \quad \quad \times H_n\left[\sqrt{\frac{\omega_0}{\hbar}}\;\left(\frac{x}{1 + \lambda x}\right)\right] dx. 
\end{eqnarray}
The above integral can be evaluated to obtain the normalization constant as 
\begin{equation}
N_n = \left(\frac{\sqrt{\frac{\omega_0}{\hbar}}}{\frac{\sqrt{\pi}}{2}\;2^n\;n!\;\left(1+erf\left(c\right)\right)+c\;e^{-c^2}O(c^2)}\right)^{1/2}, \label{nn3} 
\end{equation}
where $erf\left(\frac{1}{\lambda}\sqrt{\frac{\omega_0}{\hbar}}\right)$ is  the error function and ${\displaystyle c = \frac{1}{\lambda}\sqrt{\frac{\omega_0}{\hbar}}}$.  

\begin{figure}[!ht]
\vspace{0.5cm}
\begin{center}
\includegraphics[width=0.5\linewidth]{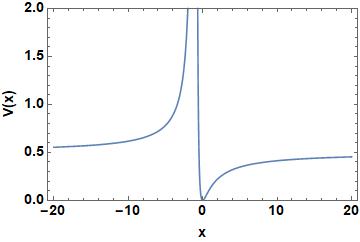}
\end{center}
\vspace{-1cm}
\caption{The plot of $V_2(x)$  for $\lambda =1$ and $\omega_0 = 2$.} \label{pot2}
\vspace{-0.3cm}
\end{figure}

\begin{figure}[!tbp]
  \centering
    \begin{minipage}[b]{0.4\textwidth}
    \includegraphics[width=\textwidth]{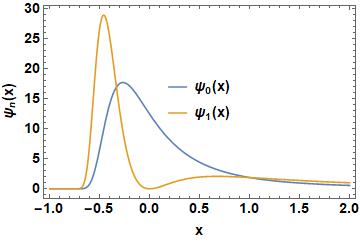}
     \end{minipage}
  \hfill
  \begin{minipage}[b]{0.4\textwidth}
    \includegraphics[width=\textwidth]{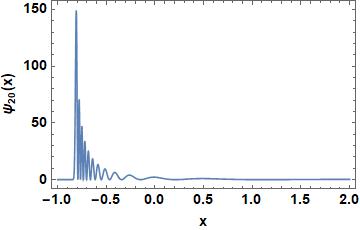}
  \end{minipage}
  \caption{The plot of $\psi_0(x)$, $\psi_1(x)$ and $\psi_{20}(x)$ for $\lambda =1$ and $\omega_0 = 2$.} \label{eig2}
\end{figure}

Figures \ref{pot2} and \ref{eig2} are plotted for positive value of $\lambda$, that is $\lambda =1.0$. Hence the potential $V_2(x)$ becomes $\infty$ at $x =- \frac{1}{\lambda}$ and hence the eigenfunctions $\psi_n(x)$ becomes zero at that point. The oscillations of the probability curves are restricted within the range $-\frac{1}{\lambda}<x< \infty.$ 
\subsubsection{Case $\lambda < 0$}
We can write down the eigenfunctions by substituting $\lambda = -|\lambda|$ in (\ref{eigen3})
\begin{equation}
\hspace{-2.5cm}\psi_n(x) = \left\{
\begin{array}{ccc}
& &\hspace{-0.6cm}\frac{N_n}{(1 -|\lambda| x)}\;\exp{\left(-\frac{\omega_0\;x^2}{2\;\hbar\;(1 - |\lambda| x)^2}\right)}\; 
H_n\left[\sqrt{\frac{\omega_0}{\hbar}}\;\left(\frac{x}{1 - |\lambda| x}\right)\right], \quad x \in \left(-\infty, \frac{1}{|\lambda|}\right),\nonumber  \\
& &\hspace{-0.6cm}0,\hspace{8.5cm}  x > \frac{1}{|\lambda|}, \label{eigen3ne}
\end{array}\right. 
\end{equation}
with energy eigenvalues, $E_n$ as
\begin{equation}
 E_n = \left(n + \frac{1}{2}\right)\;\hbar\;\omega_0, \qquad n = 0, 1, 2, 3, ...,
\end{equation}
where the normalization constant  $N_n, \; n = 0, 1, 2, 3, ...,$ is evaluated to be 
 
\begin{eqnarray}
N_n = \left(\frac{\sqrt{\frac{\omega_0}{\hbar}}}{\frac{\sqrt{\pi}}{2}\;2^n\;n!\;\left(1+erf\left(c\right)\right)+c\;e^{-c^2}O(c^2)}\right)^{1/2} \label{nn4} 
\end{eqnarray} 
where ${\displaystyle c = \frac{1}{|\lambda|}\sqrt{\frac{\omega_0}{\hbar}}}$. 

When $B+d(d-3) = 0$, the equation (\ref{geq3}) reduces to be bi-confluent equation of the form as, 
\begin{eqnarray}
\hspace{-1.5cm}\;\phi''(z) +\left[\frac{2(d-1)}{z} + 2\;\mu\;z(1 - z)\right]\phi'(z) +\left[\frac{2\;\mu\;(d-1)}{z}+\frac{2\;E}{\hbar^2\;\lambda^2} + \mu - 2\mu d  \right] \phi(z) = 0, \nonumber \\ \label{gbich4}
\end{eqnarray}
which is quasi-exactly solvable. 

One can find out the class of exactly solvable potentials by considering different ordering forms satisfying the condition (\ref{constraint2}) for the system (\ref{ham2}) also as discussed in the section $3.1$.

Hence the two nonlinear oscillators $H_1$ and $H_2$ are exactly solvable provided certain constraints on the ordering parameters are introduced. 

\section{Non-Hermitian ordering and quantum solvability}
So far we have discussed about the quantum solvability of the two nonlinear systems (\ref{ham1}) and (\ref{ham2}) by solving the generalized Schr\"{o}dinger equation corresponding to the Hermitian ordered forms of the Hamiltonians. It is also of importance to understand the quantum dynamics of the systems (\ref{ham1}) and (\ref{ham2}) for the most general ordered form (\ref{geo}). It includes both the hermitian ordering and non-hermitian ordered forms (\ref{heta}). In general, one can obtain the most general solutions for the position dependent mass quantum systems of interest by solving the generalized Schr\"{o}dinger equation  associated with the general ordered form (\ref{nhe}). In an alternate way, we can obtain the solutions of the equation (\ref{nhe}) by using its quasi-Hermitian property. As we have already seen that the non-Hermitian ordered Hamiltonian $\hat{H}_{non}$ is related with the Hermitian ordered Hamiltonian $\hat{H}_{her}$ with respect to $m^{\eta}$ as given in (\ref{heta}), we can relate the eigenfunctions of $\hat{H}_{her}$, say $\psi$,  with that of the non-Hermitian Hamiltonian $\hat{H}_{non}$, say $\hat{\phi}$, through the equation,

\begin{equation}
\hat{\phi} = m^{-\eta}\psi, \qquad \eta = \frac{\bar{\gamma}-\bar{\alpha}}{2}. 
\label{solu-meta}
\end{equation}
This is known as the quasi-Hermitian property of the Hamiltonian $\hat{H}_{non}$.

As we know the solutions of the system, $\hat{H}_1$ (vide (\ref{ham1})), resulting from (\ref{geq1})  which reads as (\ref{eigen1}), we can obtain the solutions associated with the non-Hermitian ordered form (\ref{nhe}) through (\ref{solu-meta}) for the value $\lambda > 0$ as
\begin{eqnarray}
\hspace{-2.5cm} \qquad \hat{\phi}_n(x) &=& N_n\;\lambda^{(\bar{\alpha}-\bar{\gamma})}\exp{\left[(\bar{\alpha}-\bar{\gamma})\lambda\;x\right]}\;\exp{\left(-\frac{\omega_0}{2\;\hbar}e^{2\;\lambda x} + \frac{\omega_0}{\hbar} e^{\lambda x}\right)}\; e^{\lambda x/2} \; \nonumber \\ 
\hspace{-2.5cm} \qquad & & \qquad \quad \times H_n\left[\sqrt{\frac{\omega_0}{\hbar}}\;\left(e^{\lambda\;x}-1\right)\right], \quad -\infty < x < \infty, 
\label{eigen1-non}
\end{eqnarray}
and for $\lambda < 0$, we can get 
\begin{eqnarray}
\hspace{-2.5cm} \qquad \hat{\phi}_n(x) &=& N_n\;\;|\lambda|^{(\bar{\alpha}-\bar{\gamma})}\exp{\left[-(\bar{\alpha}-\bar{\gamma})|\lambda|\;x\right]}\;\exp{\left(-\frac{\omega_0}{2\;\hbar}e^{-2\;|\lambda| x} + \frac{\omega_0}{\hbar} e^{-|\lambda| x}\right)}\; e^{-|\lambda| x/2} \; \nonumber \\ 
\hspace{-2.5cm} \qquad & & \qquad \quad \times H_n\left[\sqrt{\frac{\omega_0}{\hbar}}\;\left(e^{-|\lambda|\;x}-1\right)\right],  -\infty < x < \infty, \label{eigen2-non}
\end{eqnarray}
with energy eigenvalues $E_n$ as
\begin{equation}
E_n = \left(n + \frac{1}{2}\right)\;\hbar\;\omega_0,\qquad n = 0, 1, 2, 3, .... \label{en1-non}
\end{equation}

If the Schr\"{o}dinger equation which corresponds to hermitian ordered form of the Hamiltonians results in square integrable functions on the configuration space $\mathbb{R}$ with respect to the  measure $dx$, ${\psi} \in \mathbb{L}^2(\mathbb{R}, dx)$,  the set of eigenfunctions ${\psi}$ becomes square integrable and form a Hilbert space, ${\cal H}$.  We can then obtain the eigenfunctions corresponding to the non-Hermitian form directly through the relation (\ref{heta}), as $\hat{\phi} = m^{-\eta}\;\psi$, which are well defined in the space $\mathbb{L}^2(\mathbb{R}, d\mu)$. It means that the eigenfunctions $\psi$ are square integrable with respect to $dx$ and form the Hilbertspace ${\cal H}$, while the eigenfunctions ${\hat{\phi}}$ are square integrable with the measure $m^{2\eta} dx$ and form the Hilbert space ${\cal H}'$, which is isomorphic to the Hilbert space ${\cal H}$.  Hence, the normalization constants $N_n$ in (\ref{eigen1-non}) and (\ref{eigen2-non}) are the same as (\ref{nn1}) and (\ref{nn2}) respectively.

Similarly, we can obtain the most general solutions for the system (\ref{ham2}) for the general ordered form (\ref{nhe}) using the solutions (\ref{eigen3}) as,  
\begin{equation}
\hspace{-2.5cm} {\hat{\phi}}_n(x) = \left\{
\begin{array}{ccc}
& &\hspace{-0.6cm} N_n\;\frac{1}{(1 + \lambda x)^{2\;(\bar{\alpha}-\bar{\gamma})+1}}\;\exp{\left(-\frac{\omega_0\;x^2}{2\;\hbar\;(1 + \lambda x)^2}\right)}\; 
H_n\left[\sqrt{\frac{\omega_0}{\hbar}}\;\left(\frac{x}{1 + \lambda x}\right)\right], \quad -\frac{1}{\lambda} < x < \infty,\nonumber \\
          &  &\hspace{-0.6cm}0, \hspace{8.5cm}\quad x < -\frac{1}{\lambda}, \label{eigen3-non} 
\end{array}\right. 
\end{equation}
and the eigenfunctions for the case $\lambda < 0$ we can get from (\ref{eigen3ne}) as, 
\begin{equation}
\hspace{-2.5cm}\hat{\phi}_n(x) = \left\{
\begin{array}{ccc}
& &\hspace{-0.6cm}\frac{N_n}{(1 -|\lambda| x)^{2\;(\bar{\alpha}-\bar{\gamma})+1}}\;\exp{\left(-\frac{\omega_0\;x^2}{2\;\hbar\;(1 - |\lambda| x)^2}\right)}\; 
H_n\left[\sqrt{\frac{\omega_0}{\hbar}}\;\left(\frac{x}{1 - |\lambda| x}\right)\right], \quad x \in \left(-\infty, \frac{1}{|\lambda|}\right),\nonumber  \\
& &\hspace{-0.6cm}0,\hspace{8.5cm}  x > \frac{1}{|\lambda|}, \label{eigen3ne-non}
\end{array}\right. 
\end{equation}
with energy eigenvalues $E_n$ as
\begin{equation}
 E_n = \left(n + \frac{1}{2}\right)\;\hbar\;\omega_0, \qquad n = 0, 1, 2, 3, .... \label{en2-non}
\end{equation}

Note that the above eigenfunctions are square integrable with the measure $m^{2\eta} dx$. Hence, the normalization constants $N_n$ in (\ref{eigen3-non}) and (\ref{eigen3ne-non}) are the same as obtained in (\ref{nn3}) and (\ref{nn4}) respectively.

\section{\label{conc} Conclusion}
We considered the quantum counterpart of the two one-dimensional quadratic Li\'{e}nard type nonlinear oscillators which admit maximal (eight parameter) number of symmetry generators. They are linearizable as well as isochronic. We studied the quantum dynamics of the nonlinear oscillators by considering a general ordered position dependent mass Hamiltonian. We observed that the quantum version of these nonlinear oscillators are exactly solvable in which the ordering parameters of the mass term though arbitrary to start with, get  subjected to certain constraints. Both the quantum systems exhibit linear energy spectrum as the classical systems exhibit isochronous oscillations. We have also extended the study to other examples of quadratic Li\'{e}nard type nonlinear oscillators exhibiting isochronous oscillations at the classical level and observed that the quantum counterpart of position dependent mass systems cannot be solved exactly in general. Some details are given in Appendix B. It may be due to the presence of quadratic coordinate variable.  It will be interesting to investigate whether all linearizable quadratic Li\'enard oscillators are exactly isotonic. We hope to pursue this question further. 

\appendix
\section{Appendix: Classical dynamics of nonlinear oscillators}
In this section, we will discuss the classical dynamics of the two one dimensional nonlinear oscillators (\ref{ham1}) and (\ref{ham2}) which belong to the quadratic Li\'{e}nard type nonlinear oscillators of the form,
\begin{equation}
\ddot{x}+f(x)\dot{x}^2+g(x)=0,  \label{lie-eq}
\end{equation}
and are shown to possess maximal eight Lie point symmetries \cite{tiwari}. In the paper, the authors have also shown that such systems are isoperiodic with the harmonic oscillator and hence they can be mapped on to the linear harmonic oscillator, $\ddot{X}+\omega^2_0 X = 0$, by the transformations 
\begin{eqnarray}
X & = & h(x), \label{app1}\\
h(x)&=& h_1 \int e^{\int f(x) dx} dx + h_2,\label{app2}\\
g(x) &=& g_1 e^{-\int f(x) dx}\int e^{\int f(x) dx} dx + g_2 e^{-\int f(x) dx}.\label{app3}
\end{eqnarray}

The relations (\ref{app1})-(\ref{app3}) have been used to obtain the solutions for the systems (\ref{ham1}) and (\ref{ham2}). 

The isochronicity condition for the systems corresponding to the equation (\ref{lie-eq}) is obtained as \cite{tiwari}
\begin{equation}
g'+f g = \omega^2_0 = constant. 
\label{iso}
\end{equation}

\subsection{ Nonlinear oscillator- Exponential form}
We consider the nonlinear oscillator (\ref{ham1}) with exponential form \cite{tiwari},
\begin{eqnarray}
\hspace{2cm}H_1 &=& \frac{1}{2\; \lambda^2}\;e^{-2\;\lambda\;x} p^2 + \frac{\omega^2_0}{2}\left(1 - e^{\lambda\;x}\right)^2.\nonumber 
\end{eqnarray}
The corresponding equation of motion is, 
\begin{equation}
\ddot{x} + \lambda \dot{x}^2 + \frac{\omega^2_0}{\lambda}\left(1 - e^{-\lambda\;x}\right)=0.
\label{ham1-eq}
\end{equation}
Here, $f(x) = \lambda$ and ${\displaystyle g(x) = \frac{\omega^2_0}{\lambda}\left(1 - e^{-\lambda\;x}\right)}$ satisfy the isochronicity condition (\ref{iso}) as $\omega^2_0 = constant$. Hence, the corresponding solutions can be obtained as  
\begin{equation}
x(t) = \frac{1}{\lambda} \; \ln(1-\lambda A \sin(\omega_0 t + \delta)), \qquad 0 \leq A \le \frac{1}{\lambda}. 
\end{equation}
Here, $A$ and $\delta$ are arbitrary constants and the solution is isoperiodic with that of the linear harmonic oscillator (with $\lambda =0$). 
Periodic motion is observed in the region, $-\infty < x \leq \frac{1}{\lambda}$. Outside the region, the solution is singular. 

\subsection{Nonpolynomial momentum dependent oscillator}
We consider the second nonpolynomial nonlinear oscillator (\ref{ham2}) allowing eight parameter symmetries, 
\begin{eqnarray}
\hspace{0.7cm}H_2 = \frac{1}{2}\;(1 + \lambda x)^4 p^2 + \frac{\omega^2_0}{2}\;\frac{x^2}{(1 + \lambda x)^2},\nonumber 
\end{eqnarray}
which is governed by the equation of motion
\begin{equation}
\ddot{x}  -\frac{2 \lambda }{1+\lambda x}\dot{x}^2 +\omega^2_0 x\;(1 + \lambda x) = 0.
\label{ham2-eqn}
\end{equation}
Here, ${\displaystyle f(x) =  -\frac{2 \lambda }{1+\lambda x}}$ and $g(x) = \omega^2_0 x\;(1 + \lambda x)$ satisfying the isochronicity condition (\ref{iso})   as $\omega^2_0 = constant$ and  hence the corresponding solution can be expressed as 
\begin{equation}
x(t) = \frac{A \sin(\omega_0 t + \delta)}{1-\lambda A \sin(\omega_0 t + \delta)}, \qquad 0 \leq A < \frac{1}{\lambda}, 
\end{equation}
where $A,\; \omega_0$ and $\delta$ are constants. These solutions are again isoperiodic with frequency of oscillations exactly the same as that of the linear harmonic oscillator within the region, $-\frac{1}{\lambda} < x \leq \infty$. Outside the region, the solution becomes singular periodically. 

\section{Quantum solvability of other isochronous nonlinear oscillators}
Recently, Mustafa  \cite{omustafa} studied the isochronicity, linearizability and exact solvability of some one dimensional and $n$-dimensional position dependent mass nonlinear oscillators corresponding to (\ref{lie-eq}). In this section, we analyze the quantum solvability of some of the one dimensional nonlinear oscillators which possess isochronous solutions \cite{omustafa}. The position dependent mass nonlinear oscillators studied in \cite{omustafa} are
\begin{eqnarray}
L &=& \frac{\dot{x}^2}{2\;(1 + \lambda^2 x^2)} - \frac{\omega^2}{2 \lambda^2}\; \ln\left(\lambda x +\sqrt{1 + \lambda^2 x^2}\right)^2, \label{mass1} \\
L &=& \frac{\dot{x}^2}{2\;(1 \pm \lambda^2 x^2)^3} - \frac{\omega^2 \; x^2}{2\;(1 \pm \lambda^2 x^2)}, \label{mass2}\\
L &=& -\frac{(\lambda x - 2)^2}{8\;(\lambda x - 1)^3}\dot{x}^2 - \frac{\omega^2}{2\;(1 - \lambda x)}x^2, \label{mass3} \\
L &=& a^2 (\nu + 1)^2\;x^{2\nu} \frac{\dot{x}^2}{2} - \frac{\omega^2\; a^2}{2} x^{2 \nu + 2}, \label{mass4} \\
L &=& \frac{1}{2} e^{2 \lambda x} \dot{x}^2 - \frac{\omega^2}{2 \lambda^2}(e^{\lambda x} - 1)^2. \label{mass5}
\end{eqnarray}

Among all these nonlinear oscillators, the nonlinear oscillator corresponding to  the exponential type (\ref{mass5}) has been already studied in this paper. We now consider the second system (\ref{mass2}) in which we replace $\lambda^2$ by $\lambda$, for convenience. The position dependent mass Hamiltonian corresponding to (\ref{mass2}) is 
\begin{equation}
H = \frac{1}{2}\left[(1 + \lambda x^2)^3\; p^2 + \frac{\omega^2 x^2}{(1 + \lambda x^2)}\right]. \label{ham3}
\end{equation}

The generalized Schr\"{o}dinger equation (\ref{seg_he}) for the system (\ref{ham3}) can be written as 
\begin{eqnarray}
\hspace{-1cm} \; \frac{d^2 \psi}{d x^2} + \frac{6 \lambda x}{1 + \lambda x^2} \frac{d \psi}{d x} + \left[\frac{A \lambda}{1+ \lambda x^2} + \frac{B\;\lambda}{(1+ \lambda x^2)^2} + \frac{\xi \lambda}{(1 + \lambda x^2)^3}-\frac{\omega^2}{\hbar^2}\frac{x^2}{(1 + \lambda x^2)^4}\right] \psi = 0, \nonumber \\ \label{masseq}
\end{eqnarray}
where, 
\begin{eqnarray}
A &=& -36 \bar{\alpha \gamma}-15 (\bar{\alpha}+\bar{\gamma})-9 (\bar{\gamma}-\bar{\alpha})^2, \\
B &=& 36 \bar{\alpha \gamma}+12 (\bar{\alpha}+\bar{\gamma})+9 (\bar{\gamma}-\bar{\alpha})^2, \\
\xi &=& \frac{2 E}{\hbar^2 \lambda }. 
\end{eqnarray}
With the transformations,  ${\displaystyle \psi(x) = \exp{\left(\frac{\omega}{2\;\hbar \lambda (1 + \lambda x^2)}\right)}}\; \phi(x)$ and ${\displaystyle z = \frac{1}{1+ \lambda x^2}}$, we can transform the equation (\ref{masseq}) to be
\begin{eqnarray}
\hspace{-2cm} \; z  (1-z) \frac{d^2 \phi}{d z^2}+\left[-\frac{3}{2} + z + \frac{\omega}{\hbar \lambda}z (1-z)\right] \frac{d \phi}{d z} + \left[\frac{A}{4\;z}+\frac{B}{4}-\frac{3\omega}{4\;\hbar \lambda} + \left(\frac{\xi}{4}+\frac{\omega}{2\hbar\;\lambda}\right)z\right] \phi = 0. \nonumber \\ \label{ge3} 
\end{eqnarray}
Under the transformation, $\phi(z) = z^d S(z)$, we can reduce the equation (\ref{ge3}), with ${\displaystyle A=-4d\left(d-\frac{5}{2}\right)}$, as 
\begin{eqnarray}
\hspace{-2cm} \; z(1 - z) \frac{d^2 S}{d z^2} + \left[2 d - \frac{3}{2} + (1-2 d) z + \frac{\omega}{\hbar \lambda} z(1-z) \right]\frac{d S}{d z} + \left[\frac{B}{4}- \frac{\omega}{\hbar \lambda} \left(\frac{3}{4} -   d\right) -d(d -2) \right. \nonumber \\
\hspace{5cm}\left. + \left(\frac{\xi}{4} + \frac{\omega}{2\;\hbar \lambda} (-2 d + 1)\right)z \right] S = 0. 
\label{ge4}
\end{eqnarray} 
It is of the form of bi-confluent Heun equation which might be quasi-exactly solvable \cite{heun}.  We implement Bethe-Ansatz method, a quasi-exact treatment,  \cite{zhang, quesne2018}. Consider the differential equation of the form, 
\begin{eqnarray}
\sum^{3}_{j = 0} a_j z^j S''(z) + \sum^{2}_{j = 0} b_j z^j S'(z) + \sum^{1}_{j = 0} c_j z^j S(z) = 0, 
\label{bethe}
\end{eqnarray}
where $a_0, a_1,\;a_2, a_3, b_0, b_1,\;b_2$, $c_0$ and $c_1$ are parameters. 

Eq. (\ref{bethe}) has a $n$-degree  polynomial solution, 
\begin{equation}
S(z) = \Pi^{n}_{i = 1} (z - z_i), \qquad S(z) = 1 \quad \mbox{for}\quad n = 0, \label{st}
\end{equation}
with the distinct roots $z_1, z_2, . . . , z_n$, satisfying the Bethe-ansatz equations, 
\begin{equation}
\sum^n_{j\neq i} \frac{2}{z_i - z_j}= -\frac{b_3 z_i^3+b_2 z_i^2 + b_1 z_i + b_0}{a_4 z_i^4 + a_3 z_i^3+a_2 z_i^2 + a_1 z_i + a_0},\label{bethe-ansatz}
\end{equation}
provided the following restrictions on the parameters hold:  
\begin{eqnarray}
c_1 &=& -n b_2 - n(n-1)a_3, \label{con2}\\
-c_0 &=& (2 (n-1) a_3+ b_2) \sum^{n}_{i = 1} z_i + n\;b_1. \label{con3}
\end{eqnarray}

On comparing Eq. (\ref{ge4}) with (\ref{bethe}), we have 
$a_0= 0, a_1=1, a_2 = -1, a_3 = 0$ and  $b_0=2 d - \frac{3}{2}, b_1 = -2 d +1 +\frac{\omega}{\hbar \lambda} ,\;b_2 = -\frac{\omega}{\hbar \lambda}$ and 
$c_0 =\frac{B}{4}- \frac{\omega}{\hbar \lambda} \left(\frac{3}{4} - d\right) -d(d -2), \; c_1 = \frac{\xi}{4} +\frac{\omega}{2\;\hbar \lambda}\left(1-2d\right)$. 
The relations (\ref{con2}) and (\ref{con3}) imply that 
\begin{eqnarray}
E_n &=& \left(2 n + 2 d - 1\right)\hbar \omega. \label{en}\\
\frac{B}{4}& = & \frac{\omega}{\hbar \lambda} \left(\frac{3}{4} - d-n\right) +d(d -2) + \frac{\omega}{\hbar \lambda} \sum z_i + (2 d-1 )n. 
\end{eqnarray}
The roots $z_i's$, $i = 1, 2, 3, ...$ can be obtained through 
\begin{equation}
\sum^n_{j\neq i} \frac{2}{z_i - z_j}= -\frac{\frac{\omega}{\hbar \lambda} z_i (1-z_i) + (1-2d)\;z_i + 2d -\frac{3}{2}}{z_i (1 - z_i)}.\label{bethe-ansatz}
\end{equation}
With this knowledge, we can conclude that the isochronicity of the system (\ref{masseq}) is preserved in the energy spectrum (\ref{en}) and the energy levels are free from the nonlinear parameter $\lambda$. We mention here that the classical system (\ref{ham3}) is linearizable whereas its quantum counterpart is not exactly solvable, that is quasi exactly solvable. 

To understand the quantum dynamics of different types of isochronous nonlinear systems collectively, we also considered  the other examples (\ref{mass1}) to (\ref{mass4}) and applied  the  same  procedure as discussed above. We  observed  that  the  corresponding Schr\"{o}dinger  equations pertaining to all the three systems (\ref{mass1}), (\ref{mass2}) and (\ref{mass4}) cannot be transformed  to either  those equations  of  classical  orthogonal  polynomials  or Heun type equations. Hence, we cannot now conclude that the  systems (\ref{mass1}) to (\ref{mass4}) can be solved exactly or the systems (\ref{mass1}), (\ref{mass2}) and (\ref{mass4}) can be solved quasi exactly. Comparatively,  we can state that the classical systems may be linearizable but this does not ensure that the corresponding quantum systems can be solved exactly. In future, we plan to analyze this observation with further examples.

\section*{Acknowledgment}
VC wishes to acknowledge DST for the financial support of the project (No. SR/WOS-A/PM-64/2018(G)) under Women Scientist Scheme A. ML acknowledges the financial support under a DST-SERB Distinguished Fellowship program (Grant No. SB/DF/04/2017).

\end{document}